\documentclass[]{article}
\usepackage{graphicx}
\usepackage{caption}
\usepackage{subcaption}
\usepackage{hyperref}
\usepackage{natbib}
\title{High Entropy Alloys as a new metaphor in sociophysics}
\author{Pawel Sobkowicz}

\begin{document}

\maketitle

\begin{abstract}
Most of the opinion dynamics models in sociophysics have their historic origin in studies two-dimensional magnetization phenomena. This metaphor has proven quite useful, as it allowed to use well known techniques, relating individual behaviours of single spins and their interactions, to large scale properties, such as magnetization or magnetic domains creation. These physical properties were then ``mapped'' to social concepts: spin orientation to a person's views on a specific issue, magnetization to global opinion on the issue, etc. During the past 20 years, the models were significantly expanded, using more complex individual agent characteristics and even more complex types of the interactions, but the power of the metaphor remained unchanged.
In the current paper we propose to use a new physical system as the basis for new ideas in sociophysics. We shall argue that the concepts and tools devoted to studies of High Entropy Alloys (HEAs) could significantly broaden the range of social concepts ``addressable'' by sociophysics, by focusing on a wider range of global phenomena, arising from atomic properties, interactions and arrangements. 
We illustrate the new idea by calculating a few characteristics  of a simple HEA system and their possible ``mapping'' into social concepts. 
\end{abstract}

\section{Introduction}
The idea to use the concepts and rigour of physics to describe social phenomena dates back as far as 18th century. Marquis de Condorcet, based on his extensive knowledge of mathematics, introduced a concept of social arithmetics, ``\textit{la science qui a pour objet l'application du calcul aux sciences politiques et morales}'' in 1794. Some decades later, August Comte proposed \textit{physique sociale} (social physics) as a new science in 1839, in the fourth tome of his \textit{Cours de philosophie positive}, arguing from the point of view of promoting progress through better understanding of social phenomena, by applying to social sciences the samerequirements as to physics or chemistry. These ideas were shared by Adolphe Quetelet, who, a few years earlier (1835) published \textit{Sur l'homme et le développement de ses facultés, ou Essai de physique sociale}. His ambition was to include social sciences into the mainstream composed of mathematics, physics, chemistry and biology.

Have these early attempts at connecting mathematical foundations typical for physics with the importance of understanding our societies been successful? In certain aspects, the answer is a decisive yes. 
Most of social sciences today use highly developed statistical tools and models to derive the main characteristics of social systems and to provide mathematical strength to the reasoning or to better understand the data. 
This application of statistical methods (although in many cases imperfect due to small sample sizes) has allowed sociologists to move beyond the anecdotal and speculative ``just so stories'' stage.
On the other hand, for many research questions, the complexity of the human behaviour overwhelms the statistical and mathematical models. The latter situation is present especially when the amount of unknowns and uncertainties (typical for the complex social situations) is so large as to allow multiple, mutually incongruent models to be created and used. The appeal of the high status of physics among sociologists and psychologists (sometimes dubbed ``physics envy'') has even led to abuses, recognized as early as 1970's \citep{andreski72-1}.

In the context of our work, it is important to remember that the ubiquitous use of mathematics is not the only source of the strength of physics. Another obvious reason for its success is the constant interplay between theory and experiment, challenging and driving each other. But there is yet another aspect, often forgotten: it is the ability to pick the right level of abstraction for the studied problems. For example, in studying the physics of gases we ``forget'', for the most part, the internal complexity of atoms, their nuclei, the quark-gluon interactions -- replacing them with a simple elastic-ball approximation. Using a two parameter  Lennard-Jones potential, chosen for its mathematical simplicity leads to many important results and ideas, even though we know that this potential is not ``true''. 
Finding the right balance between the complexity of concepts used to build a model and the simplicity required for the model to be useful ind insightful is still a grand challenge of social sciences.

During the past 30--40 years, with the advances in statistical physics of complex systems and, especially, with the appearance and ubiquity of computers and computing tools, the ideas of social physics begun to be part of mainstream research. Modernized name of \textit{sociophysics} denotes today a broad range of cases where analogies with certain physical systems are used to improve understanding of social systems and human behaviours. We are now able to use these ideas to study many static and dynamic processes, sometimes with quite important and illuminating results.

The current work is mainly devoted to a branch of sociophysics (one of the oldest ones), namely the studies of opinion changes in societies and societal reactions to media, propaganda, rumours etc. The paper will start with a short introduction of the history and current status of such models, moving on to the challenges faced by the research field (Section 2). The main part is concerned with a proposal to use a relatively novel class of physical systems, High Entropy Alloys, as the base for sociophysical models. This could allow us to describe a broader range of social phenomena than the current approach (Section 3). A simple example of such model is presented in Section 4, to illustrate the potential of the approach. 

\section{Spinsons and their interactions}
The attempts to describe the dynamics of social opinions using the analogies derived from the physics of magnetic materials have a respectable history dating back more than 40 years. The foundation of such approach rests on three ``mappings'': a person is equated to an atom; the person's opinion to the orientation of the atomic spin; and the social influences are modelled by interatomic interactions. This simplification of human behaviour has led to coining of a notion of spinson (spin-person, \citet{nyczka13-1}). 

There are more correspondences in such approach. The topology of social contacts was typically one of three variants: fully connected network, in which any person could communicate with any other person; a regular grid (typically two-dimensional square lattice, the easiest one for graphical presentation of results), where the agents could interact with their neighbours (either only the closest neighbourhood or a suitably chosen extended one); a random network of a given density of links. 
 Somewhat later on, the growing popularity of network science has led to models based on more realistic interaction connections, including the scale free topology. Volatility of individual opinions corresponds, in some models to  temperature. Additionally, external influences (such as exerted by media or propaganda) were mapped to external magnetic field.

The last component of the models was the interaction mechanism -- describing the influence of one agent on others. This could take the form of one-to-one interactions (one agent trying to convince the other, or two agents sharing a discussion and trying to convince each other, for example the voter model \citep{cox86-1,bennaim96-1,galam02-4,castellano03-1}, the bounded confidence model \citep{deffuant00-1,deffuant02-1,weisbuch03-1,weisbuch03-2}) or many-to-one group influence models (where a suitably averaged opinion of a group ``surrounding'' an agent influences its opinion, for example the social impact model \citep{nowak90-1,nowak96-1,holyst01-1,kacperski00-1,kacperski99-1},  the Hegselmann-Krause model \citep{hegelsmann02-1}) or the Sznajd model \citep{sznajd00-1,stauffer01-2,stauffer02-1,stauffer02-2,slanina04-1,sabatelli03-1,sabatelli03-2,bernardes01-1}. The specific conditions required for an agent to keep or to change its opinion varied between the models, leading to different process dynamics. 
There was, however, an assumption which was used in most of the early models: any interaction between  two agents with differing opinions would diminish the difference or, at worst, leave it unchanged. 
Similarly, a group influence would tend to bring the agent's opinion closer to the group one.
This basic individual dynamics, in the absence of constraints, should eventually lead to a convergence of opinions. In fact, many of the early works considered the dynamics of the appearance of such consensus, seemingly without taking into account the observation that societies very seldom reach the state of global agreement of views.

Another characteristic feature of the spinson and related models was the use of ``high temperature state'' as the initial point. In the studies of spontaneous (or induced) magnetization, preparing the sample at high temperature results in a random orientation of spins. 
When the temperature is lowered, magnetization may appear, either globally or in domains. 
It is tacitly assumed that the specific initial random configuration is irrelevant for the final results.
Unfortunately, social systems are very seldom in such randomized state.
The evolution might crucially depend on even minor correlations within the initial state.
For this reason, the starting conditions for most models were largely unrealistic. 

The early sociophysical works were later developed to include many  constraints, inhibiting this trivial path to consensus. 
Among the most widely recognized was the bounded confidence model, 
which assumed that the opinions of the two interacting agents could get closer to each other only if the initial difference was suitably small, below the tolerance threshold. 
Depending of the tolerance level, this could lead to consensus 
(if the tolerance was high and most agents would be able to ``talk'' to each other). 
For small enough tolerances, a stable coexistence of two or more groups would appear. Each group having a locally convergent opinion, separated from other groups in a way effectively blocking intergroup communication.

It should be noted that in the standard formulation of the bounded 
confidence model, the split opinions were always within the spectrum 
postulated in the initial configuration. Thus the basic bounded confidence model could not explain
\textit{growing} polarization.
The bounded confidence model itself has been developed to cover many variants, 
some of them including repulsion of suitably different opinions, 
considerations of multidimensional opinion spaces, presence of minority groups holding fast to their opinions and many others. 
Eventually, some of these extensions have allowed to reproduce widening of gaps between social groups.

The evolution of the bounded confidence model is but one example of the changes to the initial palette of approaches. The initial interaction terms, borrowed directly from spin theory of magnetism were expanded, additional classes of agents (extremists, contrarians, inflexibles, leaders) were added. New models included dynamic networks, in which topology of links evolved in parallel to the opinions. The field is active and flourishing.

\subsection{Problems, limitations and current solutions}

One could ask why should we consider new analogy between physics and social sciences if the current approach is so active?

There are basically two answers to this question. The first is based on the observation that a certain fraction of sociophysical research stopped even pretending to be part of the multidisciplinary effort. \citet{sobkowicz09-1} has called for a closer link of models with reality (not just to prove the usefulness of the models in social contexts, but also to provide the necessary theory-experiment/observation loop). Indeed, some authors have heeded the call, and the number of papers aimed at using sociophysics to understand real social systems is growing. 
These approaches often steer away from the direct reference to physical counterparts, and are more accurately classified as Agent Based Models rather than sociophysical ones. 
The agents are in some aspects more human-like, more complex in their individual behaviours, for example incorporating emotions, biases, motivated reasoning. Also the interactions between the agents become much more complex and non-linear.

Still, there is a large number of papers which, in our opinion, use the excuse of social modelling, but should be more accurately described as ``physics of non-physical systems''. The range of properties, arrangements and interactions between actual atoms in magnetic systems is limited. Once the barrier requiring that a model should correspond to a real system is broken, one can include whatever comes to mind: multiple ``spins'' and other degrees of freedom per atom, adjustable interaction networks, multiple layers of such networks etc. 
Then one can apply the same methods, concepts and variables as are used for real physical systems and -- quite often -- derive  results not present in real physical systems for obvious reasons, and  unusual enough to warrant a publication.
If, as is often the case, there's no corresponding social system, the work represents, unfortunately, modelling for modelling's sake.

The second reason for the potential usefulness of looking for a broadened physical base of social studies analogy is more fundamental. Most of the previous work focused on the effects of putting together large numbers of agents and examining how the interactions change their behaviours. Would the individual reactions to external stimuli become ordered/coordinated? Can a small group of people convince larger groups to their views (and what are the conditions necessary for this to happen)? Can opposing views coexist -- or even grow apart, even when people have, potentially, access to the same information?

These are important questions, no doubt. But there are others, which are more difficult to address in using the current metaphor. 
They deal with behaviours of groups and societies as complex entities themselves. Focused on sociology rather than social psychology. 
Can our societies survive increased polarisation and function effectively? 
What would be the effects of technological and cultural changes? 
is the pace of such changes an important factor, and if yes, in what way?
What will be the effects of the growing strain between the super-rich and low-income populations? 
What would result from the clash of societies in a global market and  global inequalities between countries and communities?
These issues are reminiscent of looking at the macroscale properties of materials: their hardness, plasticity, corrosion resistance, melting temperatures, thermal and electrical conductance, resistance to radiation etc., etc. Of course, ultimately, these large scale characteristics have their origin in micro- and nanoscale composition and arrangement of the constituent atoms.
Our paper comes directly from the idea of such correspondence between ideas of materials science and sociology.

\section{Need for a richer metaphor}

It is not the goal of the current paper  present a ready solution or a fully developed model, describing and predicting the behaviour of some social system. \textbf{We aim at suggesting a possibility of using a different domain from physical research as the source of inspiration.}
The motivation is partially described in the previous sections: to allow creative and imaginative multidisciplinary cross-polination of ideas, and to rejuvenate sociophysics.

The inspiration for the particular choice of the of the High Entropy Alloys (HEAs) as the physical system to be used as the foundation of the model was actually the complexity of social systems. Spinson-based models (especially in their original two-dimensional versions) are obviously too simplistic to correspond to most of the real social situations. We need to include much more complex interactions and combinations of constituents, to be able to describe system long term as well as short term temporal behaviours and to focus on a richer set of global characteristics than magnetization alone.

We should note here that the magnetic system analogy is not the only one used in models of opinion dynamics. Other physical processes were considered, for example percolation theory \citep{chandra2012percolation,yang2015opinion} or kinetic models \citep{biswas2011mean,boudin2016opinion,toscani06-1,lallouache10-2}. In this context, the breadth of physical phenomena and tools developed to understand them allows to focus on different aspects of the social systems for which the models are used.

High Entropy Alloy studies are faced with similar challenges as social sciences. 
These include the presence of many chemical elements, each characterised by multiple different properties and possibilities of widely varying spatial organisation, changeable under environmental changes. The tools and concepts that could provide insights into the HEA-related problems could become useful in the social contexts.
Moreover, we already know that specific chemical compositions and preparation processes of high entropy alloys can lead to a wide variety of macroscopic characteristics. Thus the link between the nano- and micro-scales and global properties is quite fruitful in terms of designing materials capable of withstanding external forces and preserving desired functionality. At the same time, the path from nano-scale composition to the macro-scale behaviour is challenging to discover and describe.

It is our hope that an eventual ``mapping'' of various physical characteristics of HEAs to their social systems equivalents could enhance our capacity to understand, human societies, arguably the most complex systems we know of.

\subsection{High Entropy Alloys -- brief introduction}
Compared to the microscopic studies magnetic phenomena in physics, which are now over 100 years old (or even to the  sociophysical spinson-based opinion dynamics models, in their mid-forties), the field of High Entropy Alloys is very young. \citet{miracle2017critical} place its birth at 2004. \citet{gao2016high} notes that the interest in alloys composed of near-equal ratios of various elements predates these first publications by more than a decade. Nevertheless, the past sixteen years  were extremely active, mainly due to the combination of interesting science (including many unexpected results and fundamental discussions) and significant potential applications -- materials with designed properties (functional materials) can become crucial in many applications.

The literature on specific properties and applications of HEAs is already very broad. For an the introduction, in addition to the two general reviews mentioned above, recent reviews providing good starting points are \citet{pickering2016high,george2019high,george2020high}.

To present the concepts  related to the high entropy alloys, we chose to focus on these properties that could, in our opinion, provide bridges to interesting social phenomena, in the hope that such mapping would indicate that the potential offered by the HEA modelling toolset extends beyond that of the classical, spinson-based analogy.

\subsection{HEAs as social metaphor}
As noted above, it is definitely too early to describe in detail how the specific aspects of HEAs physics can be used to describe social phenomena. The list presented below offers suggestions of HEA characteristic features and their potential relevance in social modelling context. Note that there is no particular importance order in the list.
\begin{itemize}
	\item HEAs, as their name suggests, derive much of their properties and complex thermodynamics from the high entropy, especially configurational entropy. The presence of many different atomic species (elements) and multiple ways they can be arranged locally is directly reminiscent of human societies. This similarity promises that some tools and techniques developed for HEAs could be used to describe societal characteristics.
	\item In contrast to the spinson models, where the dynamics of the whole system resulted from the combination of individual changes of a single atom changeable  characteristic (its spin), the many properties of the constituents of HEAs are ``fixed'' (in the sense of atomic properties) while others are flexible (e.g. spin, type of chemical bonds). The presence and role of the fixed atomic properties suggests that the focus of research should shift from the individual to group dynamics. In fact, the interest in HEAs is in their global properties, for example mechanical ones (hardness, stress-strain properties, brittleness, ductility) or thermal or electrical properties (e.g. electric and thermal conduction). These describe the behaviour of the alloys under external influences, which would suggest similar application of the HEA based models in sociophyscis, focusing on the societal reaction to environment and external forces and influences, in particular the resilience or lack of it.
	\item One of the features of HEAs is the presence of so-called `cocktail effects' -- unexpected synergies resulting from mixtures of different atoms. These could result in exceptional structural properties, for example increased hardness, fatigue resistance, ductility etc. Such ``cocktail effects'' are obviously present in human societies, where various characteristics and roles of individual people combine to create specific groups and contribute to their characteristics.
	\item HEAs exhibit rich processing dependence: instead of the simple influence of temperature (characteristic for magnetic systems), the properties of the alloys heavily depend on the history of treatment, rate of temperature changes, mechanical treatment, irradiation etc., etc.
	Again, this shows significant similarity to the complexity and history effects in human societies.
	\item Another, related similarity, results from the possibility that specific alloys may be treated as systems away from thermodynamic equilibrium. For example, one often studies HEAs in ``as-cast'' conditions, where non-equilibrium or metastable phases are common. It leads to an interesting question whether sociophysics should treat societies as systems at equilibrium (as most of the spinson models do) or as non-equilibrium ones. HEA modelling toolset  is ready to handle these issues.
	\item Both HEAs and social systems exhibit mixtures of slow and fast phenomena, combining in complex ways. The physical examples may be provided by relatively slow atomic diffusion and fast effects of ionizing radiation. One could imagine corresponding social phenomena, for example global economic trends and associated cultural changes and local effects of specific events. The use of tools developed to understand HEAs, could be an interesting opportunity in these contexts.
	\item Mechanical studies of the behaviour under stress offer an interesting example of physical/social analogy. Application of slow, gradual changes in stress load may lead to abrupt and unpredictable changes in material structure: avalanches of dislocation movements and serrated (jerky) strain-stress curves. These processes often exhibit power-law behaviour. Their presence is similar to certain stress-relieving social phenomena, such as appearance of social protests. Moreover, the power-law dependence is quite ubiquitous in our social life \citep{newman04-2}.
	\item In addition to the complexity brought by the presence of many different elements, HEAs behaviour is further complicated by the potential co-existence of various atomic arrangements. A specific sample may crystalline or amorphous phases, the former corresponding to different structures (FCC, BCC, HCP\ldots). Depending on their preparation, the alloys may exhibit non-random arrangements: presence of single element precipitates, nanocrystals, condensates along grain boundaries, etc. This allows us to look for correspondences in social systems, differentiating organized and dis-organized social networks as consequences of the societal history (``preparation'') and  natural processes: assortative grouping and other locally non-random configurations and their effects on the overall characteristics of the society. 
	\item In certain circumstances, the multi-component alloys may exhibit spinodal decomposition, leading to the appearance of local pockets composed of single elements -- which could be compared to assortative matching and spontaneous segregation in social systems. In fact, as \citet{pickering2016high} notes, the experimental evidence indicates that only very few HEAs are stable as solid solutions. Most of them, under certain conditions decompose into multiple phases of differing chemical composition. This corresponds to observation that majority of human societies exhibit division into classes and groups, segregated by different individual characteristics.
	\item A similar example of non-uniform distribution of atoms of different kinds is called the Contrell clouds -- in which solute atoms gather around dislocations. And as dislocations represent imperfections in the ``organized'' single crystal structure, this could provide an analogy to increased assortative gathering of certain minorities where the social structure is less rigid and controlled.
\end{itemize}

As shown by these examples, the field of HEA research offers quite large number of potential ``matchings'' to social studies. 
Moreover, in the context of modelling, there are multiple ready tools which can be used in specific situations. In fact, there is a similar ``grand challenge'' facing HEA theoretical studies, as the one faced by sociology. 
This is often referred to as the ``Holy Grail of multiscale modelling'': creating a continuous set of tools and methods, spanning the range from first-principles, quantum mechanical descriptions at nanoscale level, through various approaches valid for the intermediate scales (hundreds, millions, billions of atoms) to the prediction of macroscopic behaviour of an alloy under specific treatment and conditions.
This corresponds to the challenge of social systems description. We do not have a universal understanding spanning the scale starting from psychological characteristics of individual people and their histories and environments, through description of interpersonal interactions ans small group behaviour to large scale societies.

What is especially interesting are the transitions between tools used at different scales: quantum mechanical calculations, Monte Carlo simulations, Molecular Dynamics, finite element methods. While a large number of highly developed tools and programs exists in each of these fields, construction of a consistent and usable ``modelling factory'', with proper validation and feedback, is still a challenging task for materials sciences.
This is mainly due to the difficulty of choice of the simplifications necessary to move from one ``resolution scale'' to the next one. But finding such bridging solution might suggest a path to be followed by social sciences which face similar problems related to complexity and disorder.

\section{A simple example}
To illustrate the potential of the HEA metaphor for sociophyscis we will present below some results for a simple toy model of an alloy (technically not even a proper HEA, as it is composed from just 3 elements). The goal was to show how specific properties of the alloy may depend on its preparation, and to suggest possible social ``interpretations'' for these properties.

\subsection{System description}
The system used in this work is an alloy of three metallic elements: Ag, Al and Au. This choice was motivated by several factors: atoms of the elements have significantly different masses, but their pure elemental crystal structures are the same (FCC) and show very similar lattice constants (4.09\AA\  for silver, 4.05\AA\   for aluminum and 4.08\AA\  for gold). Moreover, despite the differences in masses, the atomic radii are also quite similar (1.447\AA\  for Ag, 1.432\AA\  for Al, 144.2\AA\  for Au). This has allowed to construct an alloy in which the competition between different ``native'' crystalline structures would ba absent, as well as the size and lattice distortion effects. 

The system has been constructed using the Atomsk package \citep{hirel2015atomsk}. It consisted of a periodically repeated box with 200\AA\ $\times$ 200\AA\ $\times$ 200\AA\  dimensions, with a total of 472 805 atoms. Initially, the elements were placed in 12 randomly located, sized and crystallographically oriented domains (four for each element). The resulting  numbers of atoms of each element were: 186 601 for Ag, 135 142 for Al and 151 062 for Au. As the initial box was unphysically ``glued together'' from these domains it was not in thermodynamic equilibrium.

The process has allowed to create an input file for the popular Molecular Dynamics simulation engine LAMMPS (Large-scale Atomic/Molecular Massively Parallel Simulator, see \url{http://lammps.sandia.gov}), adopted to the use of Graphics Processing Unit (GPU) \citep{brown2011implementing,brown2012implementing,brown2013implementing,trung2015accelerating,trung2017GPU}. All subsequent simulations were run using LAMMPS. Interatomic forces were calculated using the  embedded atom method (EAM) alloy \texttt{setfl} potential files \citep{johnson1989alloy,zhou2004misfit}

The final atomic configurations were visualized and analysed (for example in the context of finding the local crystalline structure) using the OVITO package \citep{stukowski2009visualization, stukowski2012structure}.

\subsection{Polycrystal, mixed crystal and glassy samples}

\begin{figure}
	\centering
	\includegraphics[width=1.0\textwidth]{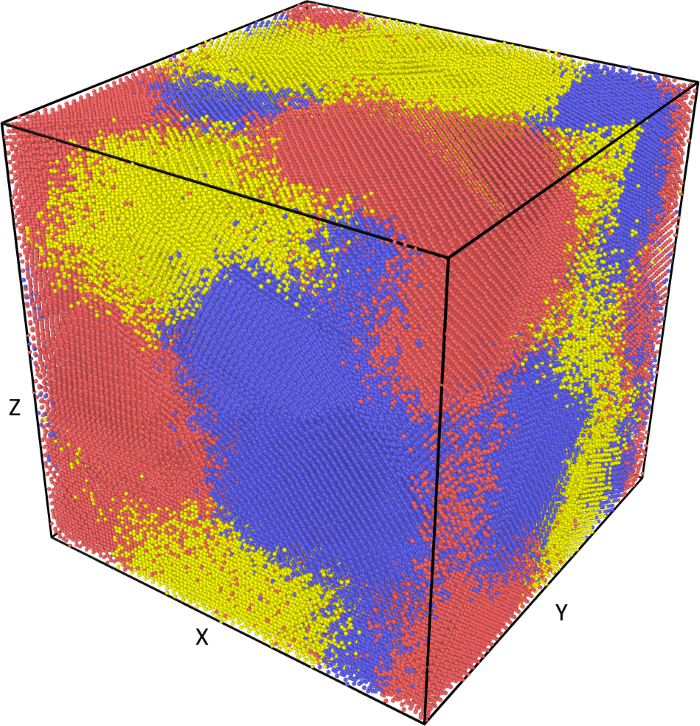}
	\caption{Three component polycrystalline box with approximately 200\AA\ $\times$200\AA\ $\times$200\AA\  dimensions (due to thermal expansion the actual size is dependent on the sample temperature), composed of slightly intermixed domains of pure elements Ag (colour red), Al (blue) and Au (yellow). The configuration results from long time equilibration at 300K of a structure artificially constructed by ``glueing together'' pure elemental grains of differing sizes and crystal (FCC structure) orientation. The elemental domains  largely preserved these differing crystal structure orientations, so in addition to boundaries between elements the sample contains  grain boundaries within the single element domains.\label{fig:polycrystal}}
\end{figure}

The initial, artificially constructed configuration, glued together from unequilibrated chunks of pure metals, was  allowed to thermodynamically equilibrate at 300K at zero pressure. This resulted in slight change of the box volume (increased by about 1\%) and some diffusion of atoms across boundaries separating different elements. The resulting multicomponent polycrystal is visualized in Figure~\ref{fig:polycrystal}.

The resulting structure was highly segregated and non-uniform. Because of the large scale of the domains, the resulting distribution differed with respect to the main axes. For example the domains composed of Ag and Al atoms formed continuous regions spanning the sample in the Z direction. For this reason we might expect that, in certain aspects, the material would behave differently with respect to directional influences along different axes. A majority of atoms were located in domains with well defined crystal structures (see Table~\ref{tab:crystalstructure}). In the following part of the paper we shall refer to this configuration as polycrystalline.

To produce the two other comparison configurations, the equilibrated polycrystal sample was heated to 2500K at zero pressure, resulting in a melted state, characterised by atomic composition disorder and lack of crystalline state. 

\begin{figure}
	\includegraphics[width=0.5\textwidth]{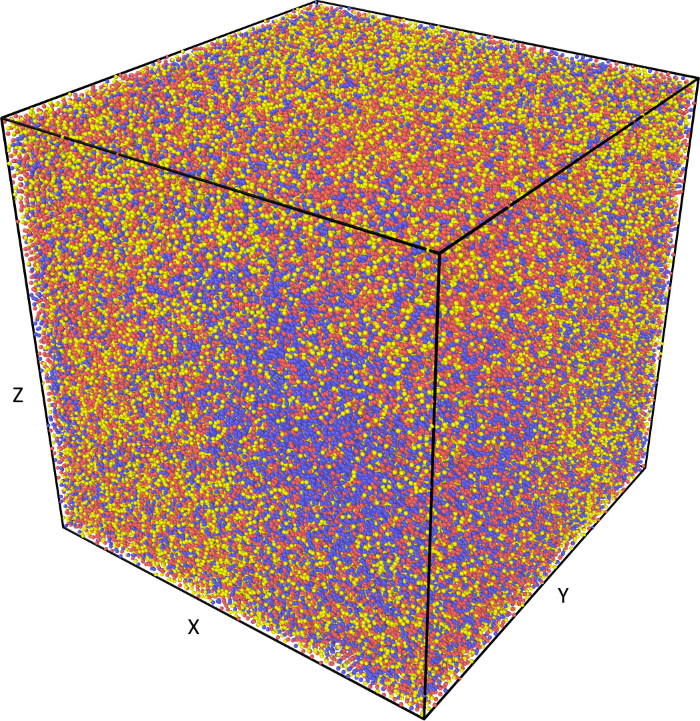}
	\includegraphics[width=0.5\textwidth]{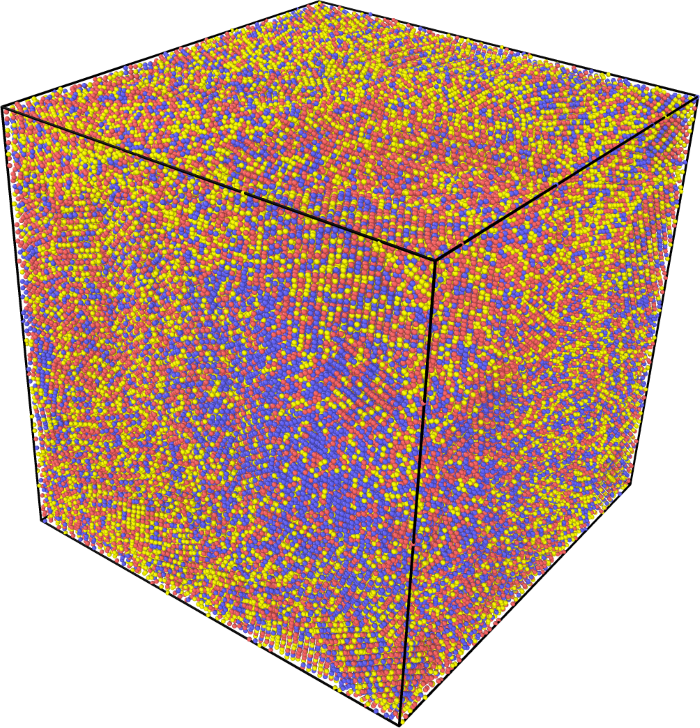}
	\caption{Comparison of the distribution of atoms of the three chemical elements between the glassy configuration (left panel) and the mixed crystal one (right panel). Glassy sample was created by melting and rapid quenching of the polycrystalline sample shown in Figure~\ref{fig:polycrystal}. Rapid cooling down resulted not only in preservation of random distribution of the elemental atoms from the melted state, but also in almost total lack of crystallized grains.
		The mixed crystal was obtained from the same melted state, but through a slow cooling, with long time spent in the 700K-500K temperature range.
		The atomic type (Ag, Al, Au) are coded by the same colours as in Figure~\ref{fig:polycrystal}. The large scale spatial distribution or atomic type for the two configurations is very similar. The difference becomes visible when one considers of the crystalline structure rather than  the chemical distribution (see Figure~\ref{fig:glassyvsmixed}). \label{fig:glassy}}
\end{figure}

The melted state was then processed in two ways. In the first, it was cooled down to 300K at a very rapid pace, resulting in creation in a quenched disorder structure, with  randomized locations of the atoms of the three elements and practically no crystal structures. Figure~\ref{fig:glassy} presents a general view of the resulting randomized simulation box. This configuration will be called glassy or quenched disorder.

In the second approach, the melted sample was allowed to cool down to 300K more slowly, with extensive time spent at 700K to 500K range. This has resulted in the atoms rearranging to form local crystalline regions -- yet still to preserve the randomness of the elemental arrangements. This configuration will be called a mixed crystal.

\begin{table}
	\begin{center}
	\begin{tabular}{|l|c|c|c|}
		\hline
		& FCC & HCP & No structure \\
		\hline
		Polycrystal & 55.7\% & 19.7\% & 24.6\% \\
		\hline
		Glassy configuration & 2\% & 1.6\% & 96.4\% \\
		\hline
		Mixed crystal & 56.3\% & 14.9\% &  28.8\% \\
		\hline
	\end{tabular}
\end{center}
	\caption{Ratio of atoms (regardless of their elemental type) classified by the OVITO analysis package as located in a specific local crystal structure, for the three initial configurations. \label{tab:crystalstructure}}
\end{table}

In a certain sense, the mixed crystal configuration may be considered to lie ``in between'' the glassy one and the polycrystal. It shares the atom type randomness with the quenched disorder system, and, at the same time, it shares the presence of locally ordered crystalline structures with the polycrystal (although we have to remember that these crystalline grains are now composed from a mixture of elements). Moreover, the mixed crystal turns out to contain several domains of different crystallographic orientations, separated by grain boundaries, resembling the domains in the original polycrystal. The sizes of these grains are somewhat smaller than in the original polycrystal sample.

The goal of  preparation of three starting configurations  was to look for meaningful mappings between the physical characteristics of the system used for MD simulations and potential social counterparts, through studies of dynamical phenomena. In particular, we hoped to observe the effects which could be mapped to different types of social ordering and diversity at a local level.

\begin{figure}
\includegraphics[width=0.5\textwidth]{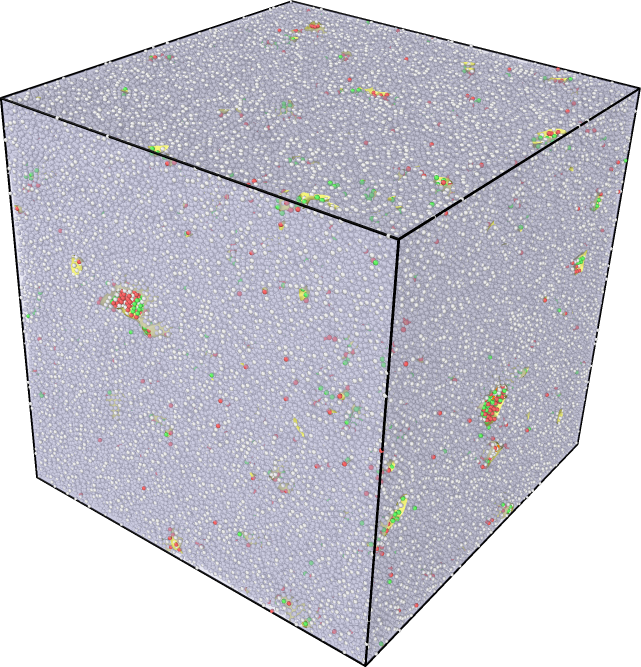}
\includegraphics[width=0.5\textwidth]{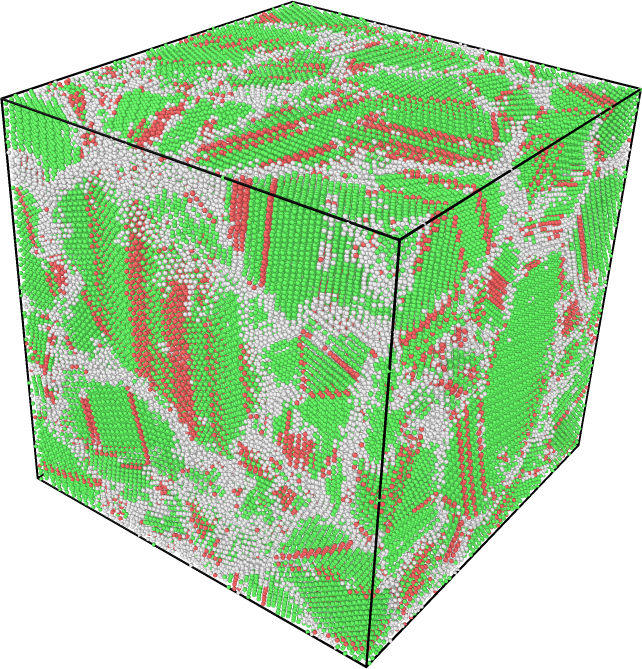}
\caption{Comparison of the local crystalline structure for the glassy (quenched disorder, left panel) and slowly cooled mixed crystal (right panel) configurations. Green: atoms identified as embedded in the FCC structure; red: HCP structure; gray: no crystal structure. The underlying atomic type distributions are very similar (see Figure~\ref{fig:glassy}), but the local spatial ordering is radically different. \label{fig:glassyvsmixed} }
\end{figure}

All configurations (polycrystalline, mixed crystal and glassy) were subsequently used to simulate system responses to various influences.  It should be noted that our choice of three metallic elements results not only in significantly different atoms' masses (and thus separate distribution of velocities at a given temperature), but also in different forms of interatomic forces. 
Indeed, the properties of the crystalline forms of pure elements are quite different: their melting temperatures of gold and aluminum differ by more than 40\%, there are significant differences in tensile strength and Young modulus etc.
In consequence, we could expect that even though the base crystalline structure for all elements is the same (FCC) and their lattice constants are similar, there would be differences between the segregated, polycrystal structure, the mixed crystal and the quenched disorder one.

In the context of social applications the goal was to show if a homogenized society (in which people with differing characteristics are not segregated but thoroughly mixed) would be significantly different from a system with dominant segregation. Moreover, one could look for a potential mapping between crystalline and glassy configurations (even for a single atomic type) and suitable social characteristics. The role of the mixed crystal would be to see if local positional structure (disregarding the atom types) could have an impact on the global properties and would have any significance in the sociophysics context. 
To provide a dynamical context we studied the behaviour of the three sample configurations with respect to two types of phenomena. The first was reaction to external forces, in a form typical for materials science, i.e. the study of the strain-stress curve. The second test was the calculation of lattice thermal conductivity. 

\subsection{Response to external forces: strain-stress curve}

The first test involves measuring the system response to external forces. The total configuration consisted of creation of ``nanowires'' by periodic repetition of the basic boxed along one of the orthogonal axes (X, Y or Z) and applying strain along this direction. 
The nanowires of (initially) square cross-section were created by repeating the basic simulation cube along the strain direction and separating it from others by adding empty space in the two perpendicular directions. 
The LAMMPS software allowed to simulate the system reaction to such strain at the level of atoms and to calculate the stress corresponding to a given strain. 

\begin{figure}
	\includegraphics[width=1.0\textwidth]{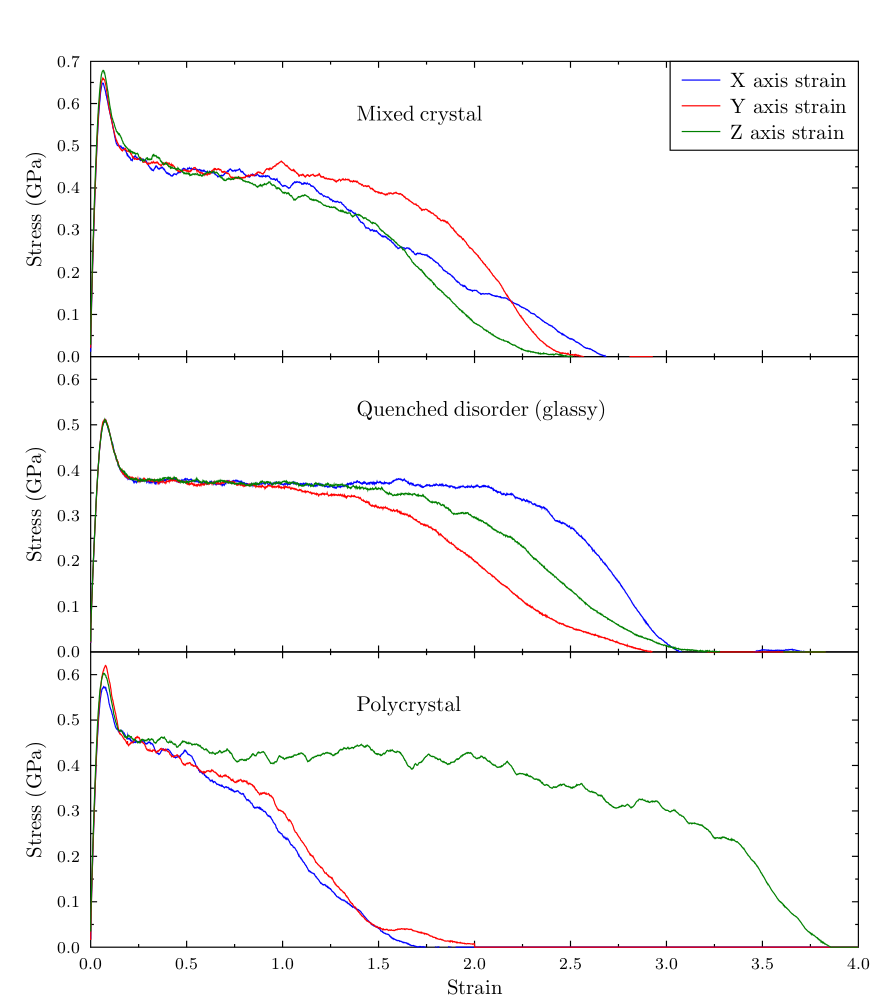}
	\caption{Stress-strain relationship for the three types of structures and strain applied along nanowires in three orthogonal directions.
	The nanowires were constructed by repeating the basic box in a given direction (X, Y or Z) and separating the boxes in the two other directions. 
	Configurations starting from high proportion of crystalline domains (polycrystal and mixed crystal) show higher values of the yield stress point, indicating higher resistance in the elastic regime. 
		They show, however, quite different behaviours in the plastic regime: there is much less directional dependence for both mixed crystal and glassy configurations, than for the original polycrystal. In the latter case, the presence of continuous domains of Ag and Al along the Z axis results in a radically greater ductility in this direction.   \label{fig:stress-strain}}
\end{figure}

Figure~\ref{fig:stress-strain} shows the resulting stress-strain curves for the polycrystalline, glassy  and mixed crystal structures. The stress-strain approach may be considered a macroscopic way of characterizing response to external forces.  For small values of strain (below 0.05) we observe a linear behaviour, typical for reversible, elastic regime. This could be interpreted in social terms as small adjustments to external drivers, which do not change the social structure. For strain above 0.1 we observe the plastic, irreversible deformation regime, in which mere reversal of strain does not return the sample to the original configuration. Here again an intuitive correspondence exists. For large enough external influences faced by a society (economic crisis, pandemic, war, internal conflict but also certain examples of technological change) the very structure of society may become changed. Specific institutions or social links may become broken and either be replaced by different ones, or vanish entirely. With increasing strain, the sample (and its social analogue) may ``stretch'', exhibiting increasing internal changes, until the point at which it would break.

\begin{figure}
	\includegraphics[width=0.9\textwidth]{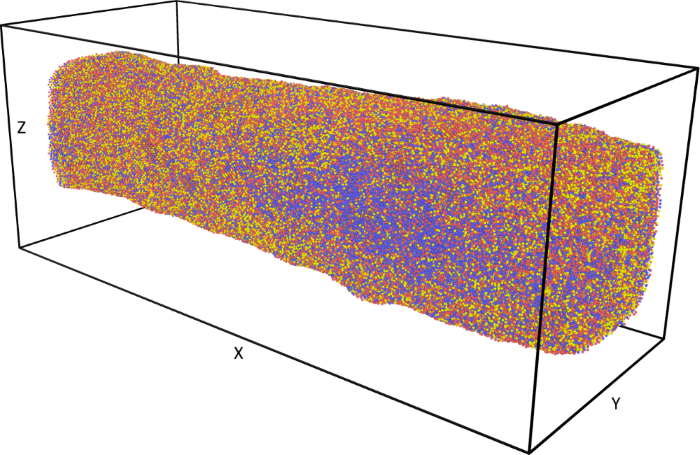}
	\includegraphics[width=0.9\textwidth]{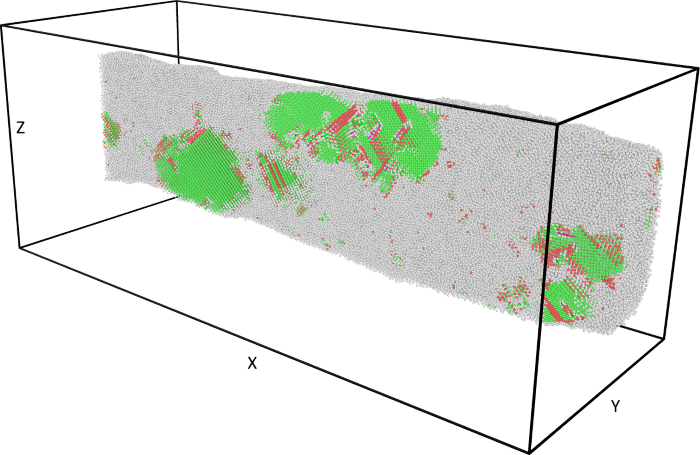}
 \caption{Microscopic view of the effect of 1.6 strain along X axis (total X direction length equal to 2.6 of the original) for the quenched disorder configuration. Top figure: overall view of atomic elements. Bottom view: slice of the sample by Y=100\AA\  plane, showing internal crystal structure. Due to strain effects, part of the sample, previously in disordered state (grey), has transitioned to crystal inclusions (green corresponds to FCC, red to HCP), composed of atoms of various types. \label{fig:strain-glassy-X}}
\end{figure}

\begin{figure}
	\includegraphics[width=0.9\textwidth]{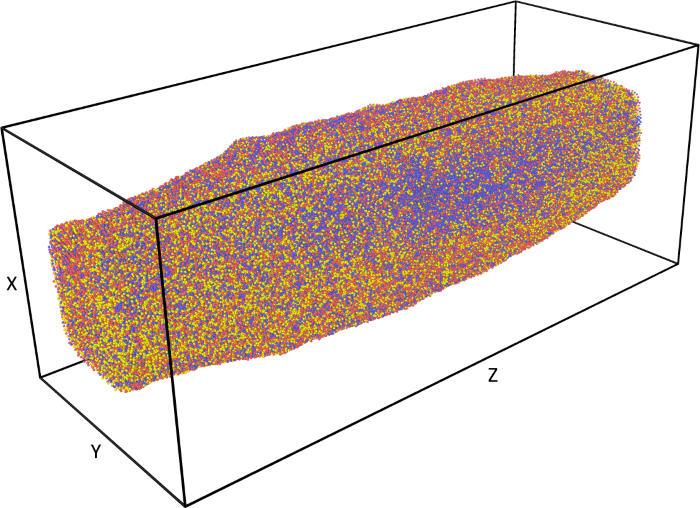}
	\includegraphics[width=0.9\textwidth]{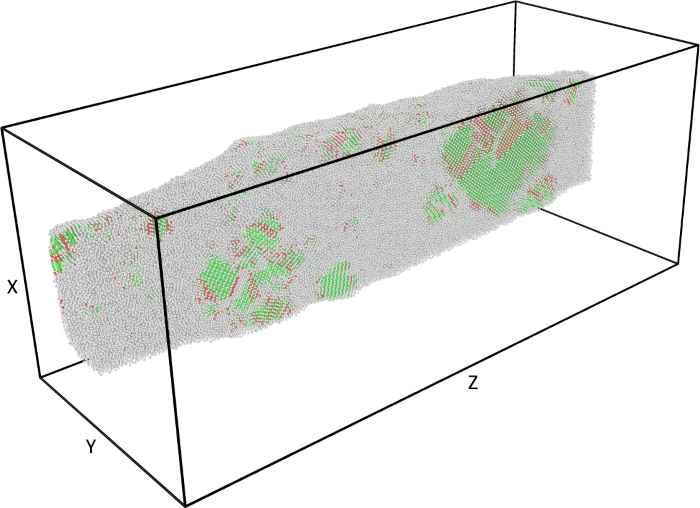}
	\caption{Microscopic view of the effect of 1.6 strain along Z axis for the quenched disorder configuration. Top figure: overall view of atomic elements. Bottom view: slice of the sample by Y=100\AA\  plane, showing internal crystal structure. Due to strain effects, part of the sample, previously in disordered state, has transitioned to crystal inclusions (green corresponds to FCC, red to HCP), composed of atoms of various types.\label{fig:strain-glassy-Z}}
\end{figure}

As expected, the differences observed for strain in different directions for the configuration with quenched disorder were relatively small. Figures~\ref{fig:strain-glassy-X} and \ref{fig:strain-glassy-Z} show  atomic composition and crystalline structure cut-outs for strain in the X and Z directions. While there are differences in specific atomic arrangements, the general behaviour is quite similar (and leads to similar strain-stress curves).

\begin{figure}
	\includegraphics[width=0.9\textwidth]{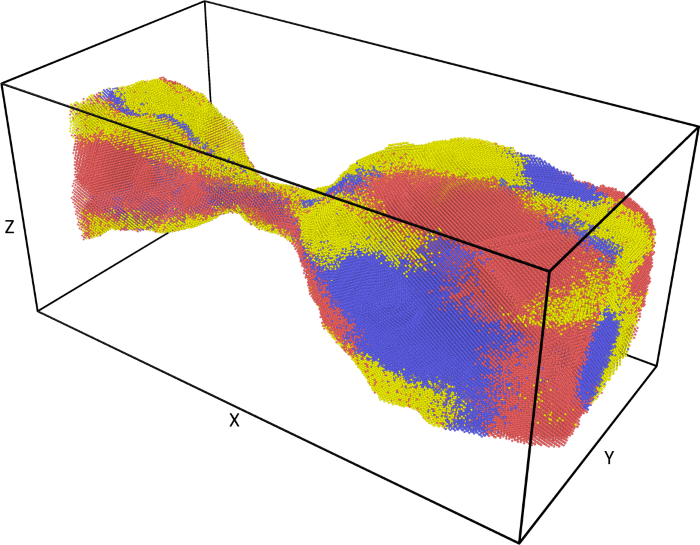}
	\includegraphics[width=0.9\textwidth]{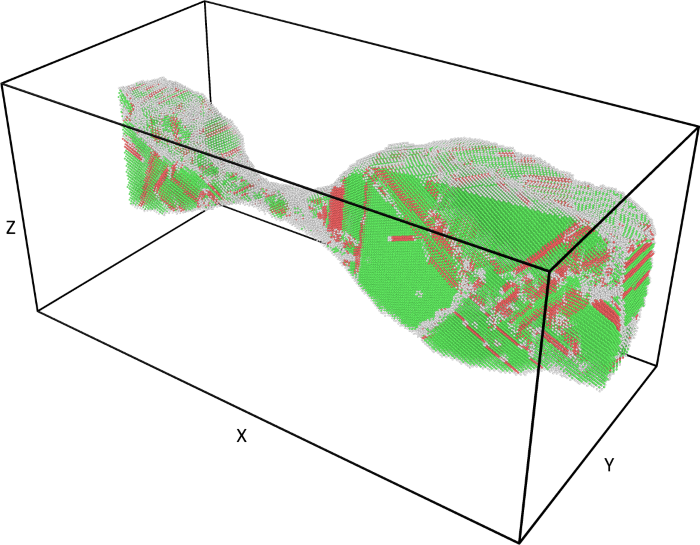}
	\caption{Microscopic view of the effect of 1.6 strain along X axis for the polycrystalline configuration. Top figure: overall view of atomic elements. Bottom view: slice of the sample by Y=100\AA\  plane, showing internal crystal structure. At this strain in the X direction, the sample is close to the breaking point. \label{fig:strain-poly-X}}
\end{figure}

\begin{figure}
	\includegraphics[width=0.9\textwidth]{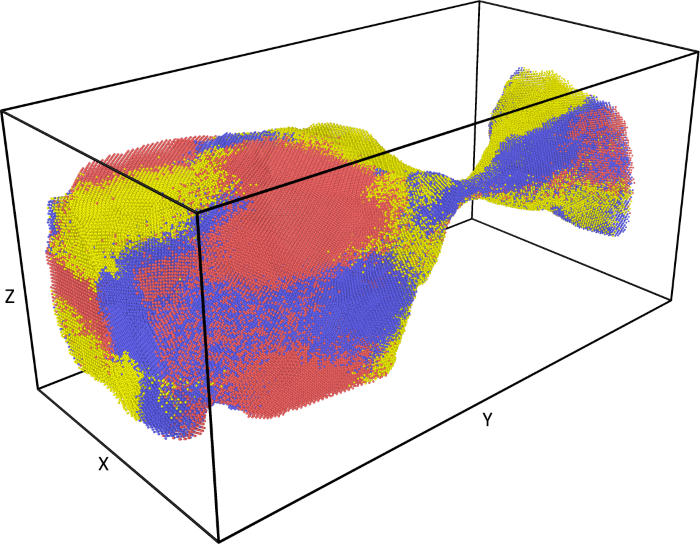}
	\includegraphics[width=0.9\textwidth]{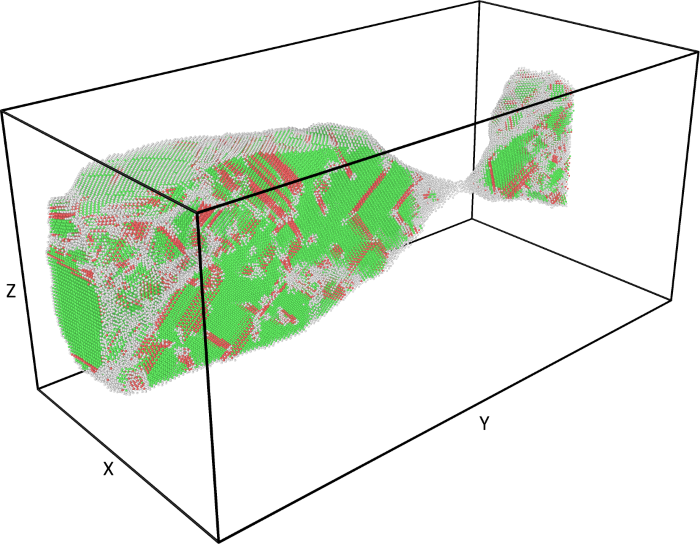}
	\caption{Microscopic view of the effect of 1.6 strain along Y axis  for the polycrystalline configuration. Top figure: overall view of atomic elements. Bottom view: slice of the sample by X=100\AA\  plane, showing internal crystal structure. As in the case of the strain in X direction (Figure~\ref{fig:strain-poly-X}) the sample is close to the breaking point.  \label{fig:strain-poly-Y}}
\end{figure}

\begin{figure}
	\includegraphics[width=0.9\textwidth]{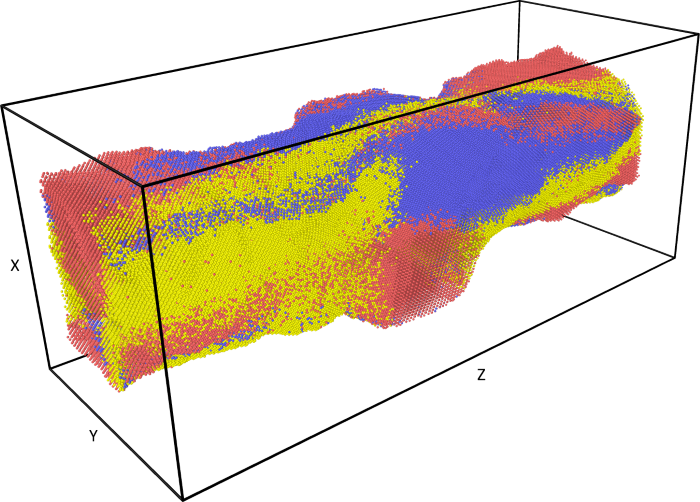}
	\includegraphics[width=0.9\textwidth]{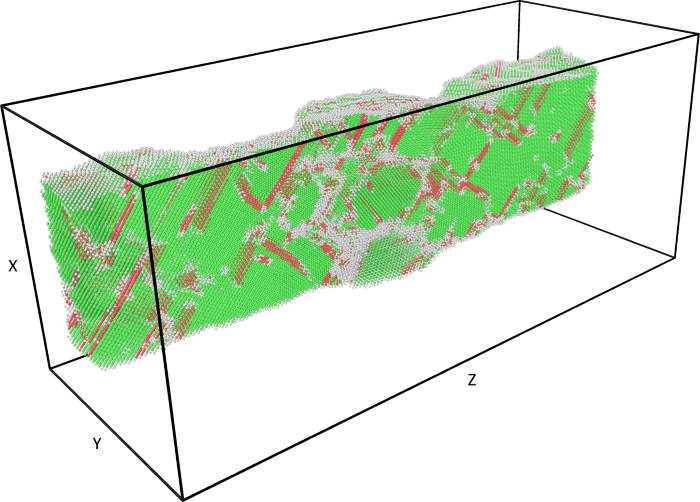}
	\caption{Microscopic view of the effect of 1.6 strain along Z axis (the same value as for the X and Y strain directions shown in the previous figures) for the quenched disorder configuration. Top figure: overall view of atomic elements. Bottom view: slice of the sample by Y=100\AA\  plane, showing internal crystal structure. Due to the presence of continuous domains of Ag and Al atoms along the Z direction of the sample, its resilience to tensile strain in this direction was much greater than for the X or Y directions. The breaking point strain is almost twice  the value for the X and Y directions (Figure~\ref{fig:stress-strain}).\label{fig:strain-poly-Z}}
\end{figure}


The Molecular Dynamics model and results described above could provide the ground for an attempt of a ``mapping'' into the language of social phenomena. We shall present such mapping in a tabular form, stressing certain similarities. However, it should be borne in mind, that the ``mapping'' is on a very crude and qualitative level, corresponding to the stage of the spinson models when spontaneous magnetic ordering below the Curie temperature was equated to social consensus.

\begin{table}
\begin{small}

\begin{tabular}{|p{0.5\textwidth}|p{0.5\textwidth}|}
	\hline
\textbf{MD model concept}	& \textbf{ Social correspondence}\\
	\hline
	\multicolumn{2}{|c|}{Concepts and characteristics}  \\ 
	\hline
Atom chemical type and the associated properties driving interactions with other atoms.	&  
Specific individual personality traits (e.g. Big Five categories: openness, conscientiousness, extraversion, agreeableness, neuroticism,  \citet{furnham2009personality}); moral foundations (e.g. dichotomies between care/harm, fairness/cheating, loyalty/betrayal, authority/subversion and sanctity/degradation, \citep{haidt2007moral,haidt2007new,graham2013moral}); behavioural patterns, education and cultural heritage -- driving individual behaviours and responses.\\
	\hline
Differences in interactions forces between pairs of atoms or larger groups, depending on atomic type, distances, spatial orientation etc.	&  
Differences in interactions between individuals of differing personality traits, moral foundations, behavioural patterns, education and cultural heritage, further complicated by the nature of their relationships. \\
	\hline
Local and mid-range spatial order (crystalline structure); partial deviation from such order (defects, dislocations, grain boundaries, twinning planes \ldots); disordered structures (e.g. glassy state or quenched disorder).	&
Presence of organized social structure, fixing ``positions'' and relationships between individuals through institutions and processes. Examples of partial deviations from fully institutionally organized society could be given by local disregard for certain laws or processes, local absence or inactivity of certain social institutions or services. The fully disordered state could correspond to societies in transition or anarchist societies, where little or no institutional social order exists.  \\
	\hline
Local order in chemical composition (for example the existence of segregation of atoms of different elements into separate volumes), precipitation of certain atoms close to crystal defects, nano-inclusions. On the opposite scale, the existence of a fully mixed solutions and alloys.	&
Existence of local communities of individuals sharing most/some of their characteristics, either by design or due to assortative matching. Societies based on segregation. Creation of local  minority enclaves. On the opposite scale: societies with high levels of diversity and no segregation.\\
\hline
External stress and sample strain; existence of elastic (reversible) deformation regime and plastic regime; notion of maximum strain (breakdown of the sample). 	&
Effects of external influences on society's or societal fragment structure (external political pressures, technological changes, environmental changes\ldots). Small and reversible responses, preserving societal structure and relative positions of individuals; irreversible changes in social structure re-modelling it to new circumstances; breakdown of society as a unified whole.	  \\
	\hline
\end{tabular}	
\end{small}	
	\caption{Mapping of concepts of the strain-stress Molecular Dynamics simulation onto the social systems \label{tab:mapping1}}
\end{table}

\begin{table}
\begin{small}	
\begin{tabular}{|p{0.5\textwidth}|p{0.5\textwidth}|}
	\hline
	\textbf{MD model result}	& \textbf{ Potential social interpretation}\\
	\hline
	\multicolumn{2}{|c|}{Simulation results}  \\
	\hline
Large difference between the breaking point strain values observed for the polycrystal (elemental segregation) and the mixed crystal and glassy samples (chemical disorder). Similarity of results for the mixed crystal and glassy samples -- low impact of crystalline order in plastic deformation regime.	&
Dominant role of diversity in response to very large social stress and changes (when the local social structure changes irreversibly).  \\
	\hline
Higher Young modulus for configurations with crystalline ordering (polycrystal and mixed crystal samples) than for the disordered case.	&
Effect of local organisational and institutional ordering, enhancing the resilience of the society to external forces below the breakdown of the institutional order (reversible regime: when external forces disappear the society can revert to the \textit{status quo ante}.)  \\
	\hline
Directional dependence in the plastic regime behaviour for the polycrystal sample, large differences in breaking point for different strain directions; small directional differences for chemically disordered samples.	&
Highly segregated societies might exhibit widely different reactions to external forces, depending on the nature of the stress (corresponding in model case to physical directions). The society might be more resilient to certain stresses thanks to the high cohesion of the segregated groups -- if these groups ``span'' the whole society. In other cases, the separation between different segregated groups might actually weaken the resilience of the society, causing breakdown much earlier. Fully diverse society would show similar resilience to a broader range of forces.  \\
	\hline
Appearance of local spatial ordering (crystalline structures) in disordered (glassy) sample under high strain.	&
May be interpreted as local self-organization   under high levels of societal change and challenges, even if these institutions and ordering were absent in the original, ``relaxed'' state. \\
	\hline
\end{tabular}	
\end{small}	
	\caption{Mapping of  results of the strain-stress Molecular Dynamics simulation onto the social systems \label{tab:mapping2}}
\end{table}


In addition to the correspondences shown in Tables~\ref{tab:mapping1} and \ref{tab:mapping2} one could also think of looking into a potential social interpretation of the microscopic concepts and phenomena occurring under stress: creation and mobility of dislocations, slip planes, twin boundaries, avalanches of local stress release and others. Other areas for model and interpretation development might cover the differences brought by macroscopic types of stress: in addition to the tension (as presented in out toy model), one could study the effects of compression, shear or torsion, leading to different types of nanoscale behaviours, and corresponding to different societal pressures and challenges.

In a similar vein, not related to the simulations presented in this paper, one could exploit the analogy between the influence of minor components of traditional alloys used to improve the characteristics of the material (such as relatively small additions to iron used to create dedicated steels) and the role of specific minorities in strengthening societies.

\subsection{Energy transmission: heat conductivity}

Another global characteristic of societies is their capacity to transmit ``something'' between remote parts through interactions between people. This ``something'' might be a material entity (various goods, or even carriers of infectious diseases) or not (rumours, ideas, innovations, emotions). Once could thus look for physical equivalents of such transmission processes, for example electrical conduction or heat transport in material samples. The analogy is limited by the fact that the range of the individual interactions between society members is not necessarily local. In this aspect, the use of Molecular Dynamics to ``map'' social phenomena where long range or one-to-many communication is dominant (e.g. information spreading) might not be suitable. 
Quantum \textit{ab initio} or Kohn-Sham calculations of electronic band structure could provide a better tool for such comprehensive calculations.
On the other hand, we could look for examples where the social transmission model is largely local in nature. In such cases the proposed approach might provide important contribution. Examples may be provided by the already mentioned epidemic transmission in local communities, spread of rumours via word-of-mouth or panic effects in crowds with visibility limited to near neighbours.  

Our choice of the toy-model alloy, while providing such advantages as the similarity of lattice constants, atomic sizes and crystal structures of the composing elements, has one disadvantage with respect to thermal conductivity. All component metals have heat transmission which is dominated by the contribution of electrons. At 300K only about 1\% of the thermal conductivity $\kappa$ is due to lattice vibrations \citep{jain2016thermal,wang2016first}. In addition, it is quite a challenge to separate the two processes, as the electron-phonon scattering may also play a role for specific materials. As our model disregards the electronic contribution, the absolute values of the thermal conductivity would be incomparable with the measured values. Instead, we'd use, as a reference, the relative phonon (lattice) components, normalized to the value for gold. \cite{jain2016thermal} provide values of $\kappa_p$ of 2.6 for Au, 5.2 for Ag and 5.8 for Al, while \cite{wang2016first} estimates the values as 2 for Au, 4 for Ag and 6 for Al. The ratios (normalized to the gold value)  are then about 2 for silver and between 2.2 and 3 for aluminum.

Before we move to the discussion of the lattice based thermal conductivity, let us note that the electronic contribution to the heat conduction may also find some analogy in social systems. Because the conduction band electrons in metals are delocalized, their contribution might be compared to long range (or even global) communication patterns (such as media). The varying relations between lattice and electronic contributions for various materials could allow modelling of various social circumstances.

The adjustment of the lattice component of the heat conduction in advanced materials is an active research subject in physics. It is  important for selecting materials with low conductivity, leading to higher resistance to radiation damage \citep{caro2015lattice} or for semiconductor materials, such as Si-Ge compounds and devices \citep{abs2013thermal,hahn2014effect,baker2015effect,lee2016thermal}. These studies have looked at the combination of effects of chemical order/disorder as well as positional ordering (crystalline, large and nano-grains and amorphous phases, but also artificially constructed arrangements, such as superlattices \citep{mizuno2015beating,wang2015optimization}). At the theoretical level, many of these works have used the same toolsets (e.g. LAMMPS codes) as our current model.

Our question in this part of the paper was quite similar to the actual : which characteristic feature of the three configuration influences thermal conduction the most? Is it the elemental separation (polycrystal) versus chemical disorder (mixed crystal and glassy state), or is it the presence of a local ordering in form of lattice structure (polycrystal and mixed crystal) vs. the quenched disorder configuration?

In contrast to \citet{caro2015lattice}, we have used the same interatomic potentials from the EAM model, rather than the simplified Lennard-Jones ones. The simulations of the heat transfer were based on the Langevin model included in the LAMMPS system.

In addition to the three alloy configurations used in the strain-stress studies, we have also used the mixed crystal geometry to provide the baseline single-element heat conductivity. To do so, we have replaced in the mixed crystal configuration all atoms with single species: Au, Ag or Al. This has allowed us to calculate thermal conductivity of pure materials, for comparison with the previously reported results.

\begin{table}
	\begin{center}
		\begin{tabular}{|p{3cm}|p{4.5cm}|p{2.8cm}|}
			\hline
			\rule[-1ex]{0pt}{2.5ex} \textbf{Material} & \textbf{Configuration} & \textbf{Thermal conductivity} $\kappa_p$, relative to gold \\
			\hline
			\rule[-1ex]{0pt}{2.5ex} Au & Mixed crystal with all atoms replaced by Au & 1 (by definition) \\
			\hline
			\rule[-1ex]{0pt}{2.5ex} Ag & Mixed crystal with all atoms replaced by Ag & 1.54 \\
			\hline
			\rule[-1ex]{0pt}{2.5ex} Al & Mixed crystal with all atoms replaced by Al & 2.18 \\
			\hline
			\rule[-1ex]{0pt}{2.5ex} AgAuAl alloy & Polycrystal, X direction & 1.20 \\
			\hline
			\rule[-1ex]{0pt}{2.5ex} AgAuAl alloy & Polycrystal, Y direction & 1.20 \\
			\hline
			\rule[-1ex]{0pt}{2.5ex} AgAuAl alloy & Polycrystal, Z direction & 1.44 \\
			\hline
			\rule[-1ex]{0pt}{2.5ex} AgAuAl alloy & Mixed crystal & 0.68 \\
			\hline
			\rule[-1ex]{0pt}{2.5ex} AgAuAl alloy & Glassy (quenched disorder) & 0.64 \\
			\hline
		\end{tabular}
	\end{center}
	\caption{Thermal conductivity of various configurations. \label{tab:thermal}}
\end{table}

In the case of pure elements we observe the expected order of the thermal conductivity values. The Al/Au and Ag/Au ratios are, however, smaller than those estimated by \citet{jain2016thermal} and \citet{wang2016first}. One of the reasons might be that even if our mixed crystal sample is made chemically uniform, it still contains differently oriented grains of a size of a few tens to a hundred Angstrom. Such structure has been already indicated as a source of decrease of phonon mean free path by \citet{hahn2014effect}. This geometric disorder is the same for all pure element configurations, which would add a scattering mechanism to all of them in a similar way.

As expected, the glassy and mixed crystal have much smaller heat conductivity than pure materials or the polycrystalline configuration. The reason is that in the two configurations the dominant chemical disorder (and resulting disorder of the masses of atoms) is present. The mass distribution disorder has been noted as the primary source of phonon scattering in SiGe \citep{hahn2014effect,baker2015effect,lee2016thermal} and in NiFe and NiCo alloys \citep{caro2015lattice}. The $\kappa_p$ values for the glassy and mixed crystal systems were independent on the direction of the heat flow. A small, but significant decrease of $\kappa_p$ in the quenched disorder case comes from additional contribution of phonon scattering due to lack of crystalline order.

The polycrystal configuration, where the atoms of various types are largely separated and the mean free path of phonons is longer, exhibits much higher values of thermal conductivity. Moreover, the conductivity in the Z direction is greater than that in X and Y directions. As in the case of greater resilience, this can be explained by the asymmetry of the initial domains distribution, and the presence of continuous domains of Ag and Al atoms in Z direction.

Assignment of social meaning to the above results is less intuitive than in the case of resistance to destructive forces. Still, if we take into account that people sharing the same characteristics (cultural heritage, political views, nationality, language) communicate more often and with greater ease and depth we could attempt to create such mapping. Phonons may be compared to messages, heat transfer to information flow; all scattering mechanisms (mass disorder, differences in forces between atoms, and positional disorder) decreasing the mean free path of the phonon to limits of local person-to-person communication, the message range; the ratios of various communication inhibitions (who speaks to whom and how often) would correspond to the hierarchy of the scattering mechanisms. A similar reasoning could be applied to communicative diseases, where a similarity of lifestyles  and more frequent contacts among people sharing the same characteristics could drive higher disease communication rates. Table~\ref{tab:mapping3} presents suggestions for the mapping between physical and social phenomena in this context.

\begin{table}[ht]
	\begin{small}	
		\begin{tabular}{|p{0.5\textwidth}|p{0.5\textwidth}|}
			\hline
			\textbf{MD model }	& \textbf{Social interpretation}\\
			\hline
			\multicolumn{2}{|c|}{Model concepts}  \\
			\hline
			Vibrational (phonon-based) heat conductivity	&
			Transmission of information vial local and small group interactions.  \\
			\hline
			\multicolumn{2}{|c|}{Model results}  \\
			\hline
			The heat conductivity for pure elements is higher than for alloys.	&
			Societies composed of agents sharing similar personality traits, moral foundations or cultural heritage have higher rates of sharing and transmission and acceptance of information, due to higher trust, confirmation bias and other psychological effects.  \\
			\hline
			Polycrystal conductivity significantly higher than for the chemically disordered cases (mixed crystal and quenched disorder). 	&
			The presence of large socially uniform domains (large scale segregation) enhances the transmission within these domains. The transmission is slowed-down only at the boundaries between different social groups, due to mistrust, echo chamber and selective attention effects.    \\
			\hline
			Little difference between conductivity in sample with crystalline order but chemical disorder and fully random glassy state.	&
			The presence of organized structures may be less important then diversity  of individual characteristics for local information transmission. (Institutional ordering might become dominant for other forms of information transmission.) \\
			\hline
		\end{tabular}	
	\end{small}	
	\caption{Mapping of  concepts and results of the heat conductivity Molecular Dynamics simulation onto the social systems \label{tab:mapping3}}
\end{table}

\section{Conclusions}

The current paper is focused on a simple idea of broadening the physical base of sociophysics applications. Whether the idea will lead to successful, creative applications remains to be seen. There are certain advantages which it offers: large number of interesting phenomena present in High Entropy Alloys offers potentially a wider range of social mappings. The fact that HEA research field is growing very fast provides a source of ideas and comparisons. Lastly, there are several well developed modelling techniques and associated software systems which can be used immediately.
In addition to Molecular Dynamics (exemplified by LAMMPS used in this work to provide an example), one could point out that the similarity of challenges faced by social studies and the physics of complex condensed matter results provides interesting opportunities.
The study of human societies spans a range from individual behaviours (including their biological and evolutionary roots, psychology, education); influences of immediate environments (family, workplace) which can be studied using small scale experimental psychology; up to sociology of large, diverse and complex social groups and whole societies. 
Not only the scope and type of relationships may vary significantly, but also the timescales of changes may span the range from ``immediate'' reactions (such as decisions based on heuristics or stereotyping, often on the scale below 1 second), through scale of hours or days for cognitive and communication functions, to lifelong evolution of both individual and social profile of a person, ultimately to historical perspective of the changes in societies. Thus, even within a single topic, one may be forced to combine a multidisciplinary portfolio of methods and tools, and to combine them into a coherent whole.

The same situation exists in studies of HEAs (in general, in materials studies). The description has to span the range from picoseconds and single atoms to years and ``samples'' (such as construction elements)  measuring many meters. 
Just as it is not possible to ``reduce'' sociology to psychology effectively, it is practically impossible to use the fundamental, \textit{ab initio} calculations for the whole scope of materials science. In materials science we use a ``ladder'' of somewhat overlapping techniques, starting from quantum mechanical calculations on the fundamental level, through density functional theory, Molecular Dynamics, kinetic Monte Carlo, continuum mechanics, constitutive equations, fracture mechanics to finite element analysis to name some of the available elements of the tool-chain. Each of these methods introduces specific concepts and simplifications, which allow it to be applied to larger structures and to look at phenomena at longer times. 

As in the psychology-sociology spectrum, the continuity of transitions between different levels of description and methods mentioned above still requires a lot of work. The choice of the simplifications leading from one ``scale'' to the next one needs to be validated. In our opinion, the fact that both disciplines undergo significant progress and face active challenges is not a weakness but a strength of the proposal presented in this work. Answers to the questions linking tools used for different system and time scales of physical systems  -- or even the acknowledgement that such links are needed -- might provide interesting insights for sociophysical applications, and, perhaps, add value to the traditional social studies as well.

Lastly, the proposed expansion of the physical basis for social studies to HEAs would provide an additional, technical advantage. The use of well developed, actively maintained open source programming, modelling and analysis tools would enhance the capacity to cross-check the model results, using standardized packages and input formats. To provide an example: the LAMMPS package used here has been recently used to study selected human behaviour related phenomena, for example urban interactions \citep{park2019nonequilibrium} or crowd movements using the social force model \citep{cornes2019fear,sticco2020re}. The key element of the idea presented in our paper is, however, not the programming toolset, but the  potential similarities between highly complex material systems, such as HEAs and certain aspects of human societies.


\end{document}